# Robo-AO: An Autonomous Laser Adaptive Optics and Science System

Christoph Baranec[*,1], Reed Riddle[1], A. N. Ramaprakash[2], Nicholas Law[3], Shriharsh Tendulkar[4], Shrinivas Kulkarni[1,4], Richard Dekany[1], Khanh Bui[1], Jack Davis[1], Jeff Zolkower[1], Jason Fucik[1], Mahesh Burse[2], Hillol Das[2], Pravin Chordia[2], Mansi Kasliwal[4], Eran Ofek[4], Timothy Morton[4], and John Johnson[4]

[1]*Caltech Optical Observatories, California Institute of Technology, Pasadena, California 91125*
[2]*Inter-University Centre for Astronomy & Astrophysics, Pune, 411007 India*
[3]*Dunlap Institute for Astronomy and Astrophysics, University of Toronto, Toronto, Ontario M5S 3H4, Canada*
[4]*Caltech Astronomy Department, California Institute of Technology, Pasadena, California 91125*
*\*baranec@astro.caltech.edu*

**Abstract:** Robo-AO, a fully autonomous, laser guide star adaptive optics and science system, is being commissioned at Palomar Observatory's 60-inch telescope. Here we discuss the instrument, scientific goals and results of initial on-sky operation.

## 1. Introduction

The next decade of astronomy is likely to be dominated by large-area surveys (see the detailed discussion in the Astro-2010 Decadal survey [1]). Ground-based transient surveys such as the Catalina Real-Time Transient Survey [2], Pan-STARRS1 [3], the Palomar Transient Factory (PTF) [4,5] and LSST, as well as space-based supernova and lensing surveys such as WFIRST, will join the GAIA all-sky astrometric survey in producing a flood of data that will enable leaps in our understanding of the universe. However, once interesting objects are found, the astronomical community requires additional observations to obtain higher-angular-resolution images, deeper images, spectra, or to follow objects at different cadences or for different periods than the main surveys.

Such a follow-up capability must be well-matched to the survey capabilities, essentially meaning that an equivalent or greater sensitivity must be provided. Given the size of planned survey telescopes, great light collecting power will be required from follow-up telescopes, along with sufficient additional observing time to cover the extraordinary number of candidate targets.

Robo-AO is the first fully autonomous laser guide star (LGS) adaptive optics (AO) and science system, specifically designed to take advantage of, and improve the sensitivity of, modest aperture telescopes. In the near-infrared, the diffraction-limited observations the system provides can give capabilities equivalent to 4m+ apertures, at much lower cost and with greater flexibility. In the visible, the system will be able to deliver modest Strehl ratios, approaching angular resolutions of 0.1 arc seconds. As an efficient robotic system, intelligently selecting from a queue of targets, it is expected to be able to observe 100+ targets per night, for the first time enabling large 1000+ target high-angular-resolution surveys.

The Robo-AO system has been designed to be both cost effective and modular; only small design changes will be necessary to port the system to different telescope architectures. The first Robo-AO system will be deployed at Palomar Observatory's 60-inch telescope (P60; [6]) during 2011 and will initially be used to execute a number of unique high-angular-resolution target surveys as well as trigger off of PTF target-of-opportunity events. A copy of the first Robo-AO system will then be developed and fielded at IUCAA Girawali Observatory's 2-m telescope in Pune, India. A third natural-guide-star-only variant of Robo-AO is currently in development for Pomona College's 1-m telescope at Table Mountain, CA, and will be used in supporting undergraduate education [7]. After a proprietary period, the design and software for Robo-AO will be made public under a General Public License. It is hoped that an adoption of the Robo-AO system on telescopes around the world will be used to exploit the flood of data produced with future top-priority wide-field surveys.

## 2. Instrument overview

Robo-AO is a highly-efficient LGS AO system with visible and infrared imaging cameras. The system comprises three main parts: a laser guide star, an integrated AO and science camera system and a robotic control system.

Robo-AO's laser guide star is a 12-W, λ=355 nm, pulsed Rayleigh-beacon focused at a line-of-sight distance of 10 km, with a range-gate of 650 m. Because the laser is invisible to the naked eye, it has been approved for use without human spotters by the Federal Aviation Authority; however coordination with US Strategic Command is still necessary to avoid illuminating critical space assets. The 69 kg laser projector box (Fig. 1; center) is mounted to the side of the telescope tube with a 25 kg adjustable pointing adapter platform. A periscope is mounted at the end of the telescope to jog the beam coincident with the telescope axis. The laser was first tested in September 2010 and was permanently installed as of January 2011. The measured Rayleigh scattering return matched the theoretical return expected from the Lidar equation [8] to within measurement error, $m_U \approx 8$, and the measured spot width of 2.1 arc seconds was expected given ultraviolet natural seeing of ~1.5 arc seconds.

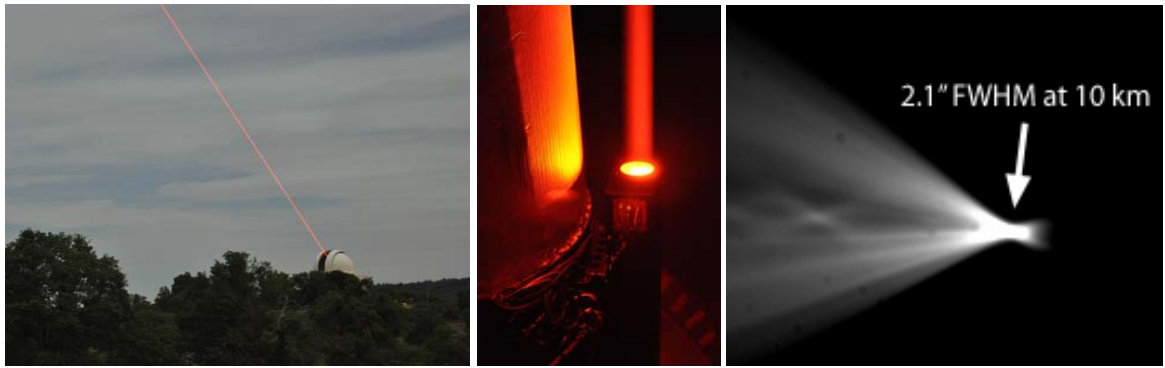

Fig. 1. The UV laser seen with a specially modified SLR camera propagating from the P60 telescope in September 2010 (left), a picture of the compact laser projector mounted to the side of the telescope tube (beam is 6.5" wide; center) and the off-axis image of the laser beacon through the P60's wide field visible camera [3] (before installation of the periscope; right.) Note that the telescope was refocused to 10 km before taking this image.

The adaptive optics system and science cameras all reside in a Cassegrain mounted, 67 kg, 1m x 1m x 0.2 m structure. High-order wavefront sensing is performed with an 11x11 Shack-Hartmann sensor, optimized for high-quantum efficiency at the laser wavelength, while high-order wavefront correction is provided by a 12 x 12 actuator micro-electromechanical systems mirror. The system currently incorporates both an electron-multiplying low-noise visible CCD and an InGaAs infrared array camera for imaging. Stellar image motion is sensed by one of the cameras on either the science object or off-axis tip-tilt guide star, and corrected by a fast tip-tilt mirror, while the other camera executes a series of scientific observations. The AO system first locked the AO control-loop at the required 1.2 kHz in the lab during December 2009, and has since been operated successfully at 1.5 kHz. Fig. 2 shows both the laboratory and on-sky versions of the AO system and science instruments. The on-sky instrument will also be able to accommodate other scientific instruments such as other cameras or fibers for fiber-fed spectrographs with appropriate additional beam splitter or dichroic mirrors.

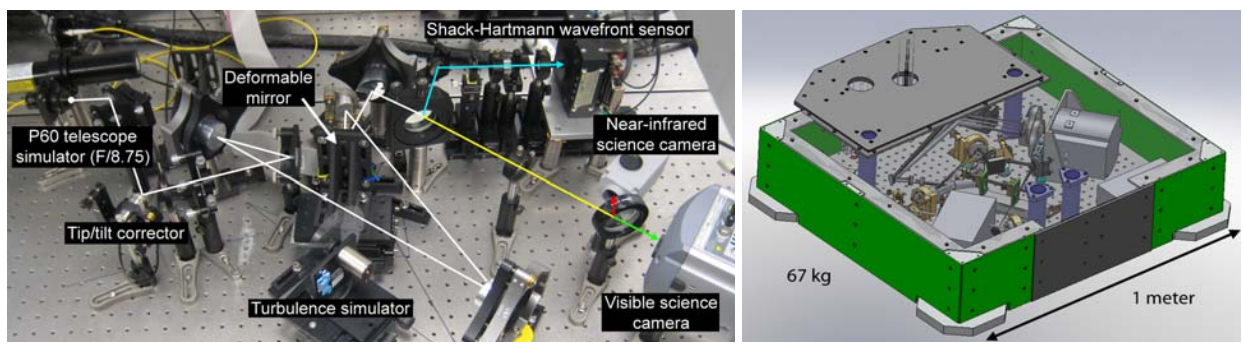

Fig. 2. The AO system and science instruments in the laboratory used for software development (left) and a CAD model of the on-sky instrument (exp. March 2011; right.)

The robotic control system operates on a consumer grade personal computer running Fedora Linux. The software is coded in C++ and incorporates extensive multi-threading and exception handling. Sub-systems, such as the core high-speed AO correction engine, operate as process daemons, subject to control by supervisor controls and oversight by watchdog processes. The robotic system operates all of the sub-systems as a single instrument. The system will be able to execute fully autonomous observations, which will be directed by an intelligent queue scheduling system.

## 3. Scientific Goals

Robo-AO has a broad range of potential contributions to astronomy, principally stemming from its ability to obtain diffraction-limited observations of very large numbers of targets. Robo-AO is designed for minimal observation overheads, including rapid automated LGS observation setups, and as such is capable of observing at least 100 targets per night. Over multiple-week surveys, which can be effectively performed by the 2m-class telescopes Robo-AO is targeted for, thousands or even tens of thousands of targets can be observed.

The Robo-AO team is planning to apply this capability in three broad areas:

**1. Large single-image surveys.** Covering several thousand targets each, these high-angular-resolution imaging surveys would be extremely time-intensive on currently available LGS AO systems. Potential projects include very large stellar binarity surveys, searches for lensed quasars [9], and follow-up observations for planetary transit searches [10].

**2. Rapid transient characterization.** New transient-search programs such as the Catalina Real-Time Transient Survey [2], Pan-STARRS1 [3] and the Palomar Transient Factory [4,5] are producing very large numbers of supernova and other transient candidates. Robo-AO will provide high-angular-resolution images of transient events within a few minutes of their detection, enabling the rapid separation of the transient from (for example) light from its host galaxy. Robo-AO's high-resolution imaging will also greatly reduce the required integration-time for infrared photometry of these faint objects.

**3. Time-domain astronomy.** Robo-AO's robotic queued operation supports recurrent, regularly spaced observations of specific targets. This will enable synoptic monitoring programs that are difficult to pursue on existing AO systems, such as long-term, high-precision astrometric characterization of binaries and searches for planets [11].

## 4. Current work

Commissioning of the entire Robo-AO system on the P60 telescope will commence in March 2011. Total commissioning time includes 35 nights over 5 separate telescope runs through July 2011. Early work will focus on operation and performance characterization of the adaptive optics system. Later work will focus on reliability and autonomous operations, culminating in the ability to execute a queue of observations during the night with no human intervention. Following the successful demonstration of autonomous operations, we will execute a month duration science campaign, showcasing Robo-AO's capabilities with a mix of large to extremely large scale high-angular-resolution surveys.

We will present the current status and recent developments from the Robo-AO deployment at the P60.

The authors gratefully acknowledge the support of the Robo-AO partner institutions, the California Institute of Technology and the Inter-University Centre for Astronomy and Astrophysics, India. This work has been supported by the National Science Foundation under grants AST-0906060 and AST-0960343.